\documentclass[draft]{agujournal2019}
\usepackage{url} 
\usepackage{lineno}
\usepackage[inline]{trackchanges} 
\usepackage{soul}
\usepackage{tablefootnote}
\usepackage{colortbl,dcolumn}
\newcommand{\GG}[1]{}
\draftfalse

\journalname{Geophysical Research Letters}

\begin{document}

%
%



\title{Out-of-plane Parallel Current in the Diffusion Regions: The Interaction Between Diffusion Region Systems and their Impact on the Outer EDR}

%
%



\authors{J. M. H. Beedle\affil{1,2,3}, D. J. Gershman\affil{2}, V. M. Uritsky\affil{2,3}, J. R. Shuster\affil{1}, T. D. Phan\affil{4}, B. L. Giles\affil{2},K. J. Genestreti\affil{5}, R. B. Torbert\affil{1}}


\affiliation{1}{Space Science Center, University of New Hampshire, Durham, NH, USA}
\affiliation{2}{NASA Goddard Space Flight Center, Greenbelt, MD, USA}
\affiliation{3}{Department of Physics, The Catholic University of America, Washington D.C. USA}
\affiliation{4}{Space Sciences Laboratory, University of California, Berkeley, CA, USA}
\affiliation{5}{Earth Oceans and Space, Southwest Research Institute, Durham, NH, USA}





\correspondingauthor{Jason M. H. Beedle}{jason.beedle@unh.edu}




\begin{keypoints}



\item Out-of-plane parallel current signatures are investigated in 26 asymmetric zero-guide EDR events and PIC simulations. 
\item These out-of-plane parallel currents are identified as defining features of the interaction between the outer EDR and IDR.
\item This parallel current is a primary signature of the outer EDR and accounts for a significant portion of the parallel energy dissipation. 

\end{keypoints}

%
%

%
%


\begin{abstract}
Dayside magnetic reconnection allows for the transfer of the solar wind’s energy into Earth’s magnetosphere. This process takes place in electron diffusion regions (EDRs) embedded in ion diffusion regions (IDRs), which form in the magnetopause boundary’s current sheet. A significant out-of-plane parallel current contribution in the diffusion regions was reported in \citeA{Beedle2023}. In order to understand the underlying structure of this parallel current, we compared EDR statistical results with a 2.5D Particle-In Cell (PIC) simulation. From this comparison, we identified out-of-plane parallel current signatures as defining features of the outer EDR and IDR. This significant out-of-plane parallel current indicates the interaction of the IDR and EDR systems, and provides implications for not only understanding energy dissipation in the diffusion regions, but also determining the location of the outer EDR. 
\end{abstract}

\section*{Plain Language Summary}

Magnetic reconnection allows for the transfer and release of energy previously stored in a magnetic field configuration. This process can occur in the boundary layer between Earth's magnetic field and the solar wind, called the magnetopause, and allows the magnetopause to act as the primary ``entry gate" for the solar wind's energy into the near-Earth environment. In the magnetopause, magnetic reconnection is triggered inside diffusion regions: an electron diffusion region (EDR) embedded inside a larger ion diffusion region (IDR). Inside these diffusion regions, plasma decouples from the magnetic field, allowing the release of magnetic energy through dissipation into the plasma, as well as creating regions of out-of-plane magnetic field, in the IDR, and out-of-plane current, in the EDR. Where this energy dissipation occurs then further divides the EDR into a central and outer EDR. From this study, we found that, through the overlap of the IDR's out-of-plane magnetic field and EDR's out-of-plane reconnection current sheet, an out-of-plane parallel current is generated that defines an interaction region between these systems. This out-of-plane parallel current is also found to act as a hallmark of the outer EDR, and allows us to better understand the generation of energy dissipation in this region.  

\newpage

\section{Introduction}

The magnetopause and its current system provide an ``entry gate" for the solar wind's energy to transfer into the magnetosphere. This process is triggered by magnetic reconnection in diffusion regions where a smaller electron diffusion region (EDR) is embedded inside the larger ion diffusion region (IDR) \cite{TreumannBaumjohann2013,HesseCassak2020}. The IDR is where ions first decouple from the magnetic field, and is defined by the Hall current and magnetic field systems driven by the still frozen-in electrons, with the Hall magnetic field taking on an out-of-plane quadrupolar, but asymmetric in strength, configuration under the magnetopause's asymmetric plasma boundary conditions \cite{Sonnerup1979,Zhang2016,Zhang2017,Jiang2022}. The EDR is then where the electrons also decouple from the magnetic field, enabling the plasma to become disassociated from the magnetic field and the reconnection process to take place -- e.g. \cite{TreumannBaumjohann2013}. This region is defined by its strong, out-of-plane electron-driven reconnection current sheet and filamentary, small-scale dissipation characteristics which are often captured by the frame-independent dissipation measure $\textbf{J} \cdot (\textbf{E} + \textbf{V} \times \textbf{B})$ -- e.g. \cite{Zenitani2011,Burch2016,Shuster2017,Shuster2021}. 

The EDR can further be split into two distinct regions, an inner EDR and an outer EDR, with the inner EDR representing the bulk of magnetic energy transfer, while the outer EDR then links this energy conversion region with the outflow exhaust \cite{Jiang2019,Huang2021,Xiong2022,Xiong2023}. Because of this distinction, the inner EDR is where positive energy dissipation is measured, indicating the conversion of the magnetic energy into the plasma, while the outer EDR is marked by negative energy dissipation \cite{Xiong2023}. 

In \citeA{Beedle2023}, a significant out-of-plane parallel current signature was found in MMS current density data from 26 dayside, no-to-low guide-field EDR events compiled from \citeA{Webster2018}. 
In this paper, we compare statistical results from these 26 EDR events with a 2.5D asymmetric PIC simulation of dayside magnetic reconnection in order to clarify where the out-of-plane parallel current is generated. In doing so, we identified this parallel current signature as a hallmark of the outer EDR and its interaction with the IDR systems.

\section{Observations}

\subsection{MMS Data}

We utilized data from MMS’s Fast Plasma Investigation (FPI) \cite{FPI} and Fluxgate Magnetometer \cite{FGM} instruments, which allowed us to consider observations of the plasma properties and magnetic field conditions from the four MMS spacecraft. The resulting plasma and magnetic field data as well as the calculated currents, were interpolated to the 30 ms FPI electron time resolution from the 150 ms ion, and the 10 ms magnetometer time resolutions respectively.

The curlometer method \cite{Dunlop1988}, was utilized to calculate the total current in the magnetopause current sheet and diffusion regions from the four MMS spacecraft's magnetometer observations by approximating gradients in the magnetic field: 

\begin{equation}
    J_{curl} = \frac{\nabla \times \textbf{B}}{\mu_0}
\end{equation}

\noindent where B is the magnetic field and $\mu_0$ is the permeability of free space. 

Utilizing $J_{curl}$, we were also able to define the components parallel and perpendicular to the magnetic field: 

\begin{equation}
	{\textbf{J}_{curl \parallel} = \left( \frac{\textbf{B} \cdot \textbf{J}_{curl}}{|\textbf{B}|} \right) \hat{\textbf{B}} \ \ , \ \  
	\textbf{J}_{curl \perp} = \textbf{J}_{curl} - \textbf{J}_{curl \parallel}}.
\end{equation}

\noindent Where, in order to match the curlometer measurements from all four spacecraft, the magnetic field data B was averaged across all the MMS spacecraft's observations. 

In addition to the current measurements, we also looked at the out-of-plane parallel current's contribution to the total out-of-plane current. In the global coordinate system utilized in \citeA{Beedle2022,Beedle2023}, the out-of-plane current corresponds to the $\phi$-direction current density where this contribution was assessed using the following expression: 

\begin{equation}
    \% J_{\phi \parallel} = \left( \frac{|J_{\phi \parallel}|}{|J_{\phi \parallel}| + |J_{\phi \perp}|} \right) \cdot 100 \%
    \label{eq:percent_para}.
\end{equation}

\noindent $|J_{\phi \parallel}|$ is the absolute value of the $\phi$ component of the parallel current density and $|J_{\phi \perp}|$ is the absolute value of the $\phi$ component of the perpendicular current density. Note, $|J_{\phi \parallel}| + |J_{\phi \perp}|$ gives the total current density that constitutes $J_{\phi \ curl}$, and allows this percent measure to work for cases where the $\phi$ parallel and perpendicular components have opposite signs. 

\begin{figure}[htbp]
    \centering
    \includegraphics[width=0.9\textwidth]{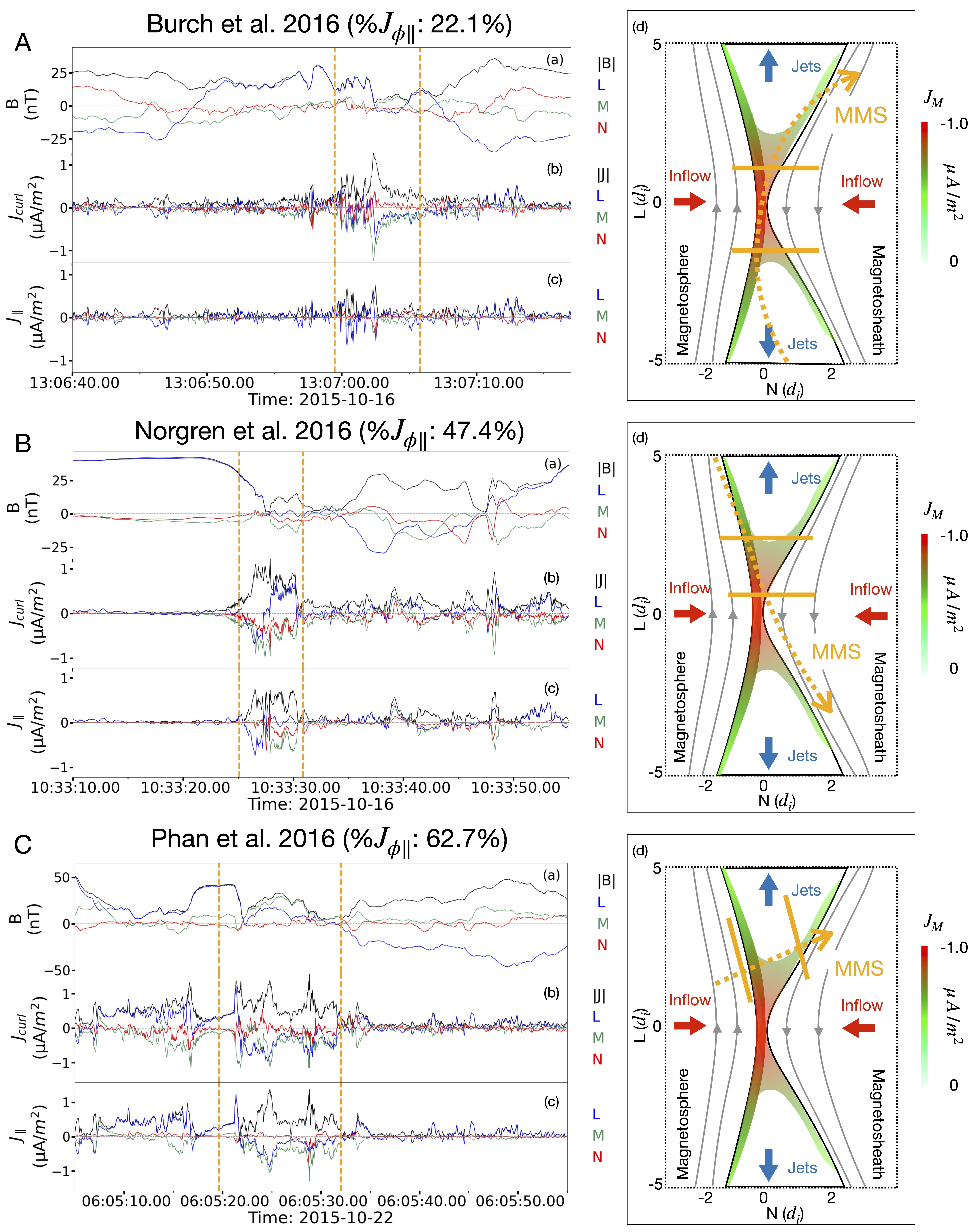}
    \caption{Combined figure depicting the time series and spacecraft trajectories through the diffusion regions of the three highlighted events in Table \ref{tab:Parallel_Table} - \citeA{Burch2016} in panel A, \citeA{Norgren2016} in panel B, and \citeA{Phan2016} in panel C. Subpanels (a)-(d) depict the MMS magnetic field data (a), current density (b), parallel current density (c), and estimates of MMS's trajectory through the diffusion regions (d) derived from the respective papers \cite{Burch2016,Norgren2016,Phan2016}. Note, the orange dashed lines in subpanels (a)-(c) represent the locations of that event's magnetopause crossing, with the approximate location of these time intervals indicated in their corresponding trajectory figures by the solid orange lines in subpanel (d).}
    \label{fig:EDR_Events}
\end{figure}

We measured the aforementioned currents and used them to construct the percent parallel quantity over 26 EDR events from the \citeA{Webster2018} study, which were previously selected and analysed in the \citeA{Beedle2023} study. The percent parallel quantity was then averaged over each of the 26 EDR crossing's magnetopause boundaries, or between the orange lines for the examples shown in Figure \ref{fig:EDR_Events}. This, along with the percent parallel's standard error ($\sqrt{\sigma}/N$) \cite{Beedle2023} and the averages for $J_{\phi \ curl}$ and $J_{\phi \parallel}$, are included in Table \ref{tab:Parallel_Table}. The table is ordered from the highest average percent parallel to least, with each event listed by its \citeA{Webster2018} event number as well as the EDR paper it first appeared in. 
Regarding differences in the events themselves, the separation among the MMS spacecraft is between 8-15 km for the majority of events, but is significantly larger for events 01, 11, and 12 --- 71.6 km, 41.8 km, and 40.8 km respectively. 

Looking at the results from Table \ref{tab:Parallel_Table}, a wide range of parallel current contributions over the 26 events can be seen. Event 24 sees the highest contribution across its magnetopause crossing at an average of over $66\%$, while event 32 sees the lowest at approximately $15\%$. Spanning this gap, the rest of the events fall somewhere between $60\%$ and $20\%$ with an average of $43\%$. Thus, even amongst this specific set of EDR events, there is a significant spread of out-of-plane parallel current contribution. 

\newpage

\begin{table}
    \caption{Table of the averaged currents and parallel current contribution over the 26 EDR events' magnetopause crossings, using the same magnetopause intervals for each event as were identified and utilized in \citeA{Beedle2023}. Each event's identifying paper is also included, and is marked by a “-" if first mentioned in \citeA{Webster2018}. Each event's \citeA{Webster2018} ID number is also included as an identifier. }
    \centering
    \begin{tabular}{ c c c c c c } 
    \\ 
    \hline
    Event Paper & Event ID & ${J}_{\phi \ curl}$ $(nA/m^2)$ & ${J}_{\phi \ curl_\parallel}$ $(nA/m^2)$ & \% ${J}_{\phi \ curl_\parallel}$ & Stan Err \\ 
    \hline
    - & 24 & 0.21 & 0.18 & 66.4\% & ±2.0\% \\ 
    \citeA{Fuselier2017} & 6 & 0.15 & 0.12 & 65.8\% & ±3.7\%  \\ 
    \citeA{Chen2017} \tablefootnote{Also included in \citeA{Ergun2017,Graham2017}} & 10 & 0.41 & 0.25 & 65.2\% & ±2.1\% \\ 
    \rowcolor{lightgray} \citeA{Phan2016} & 4 & 0.35 & 0.32 & 62.7\% & ±1.6\% \\
    - & 18 & 0.38 & 0.20 & 57.8\% & ±4.0\% \\
    - & 23 & -0.32 & -0.18 & 57.0\% & ±4.1\% \\
    - & 29 & 0.20 & 0.15 & 56.6\% & ±4.4\% \\
    \citeA{Fuselier2017} & 12 & 0.39 & 0.20 & 53.3\% & ±2.0\% \\
    - & 21 & 0.58 & 0.29 & 49.4\% & ±1.3\% \\
    - & 20 & 0.28 & 0.15 & 49.3\% & ±2.5\% \\
    \citeA{BurchPhan2016} & 8 & 0.51 & 0.29 & 48.1\% & ±5.1\% \\
    \rowcolor{lightgray} \citeA{Norgren2016} & 2 & 0.44 & 0.22 & 47.4\% & ±2.9\% \\
    - & 19 & 0.43 & 0.23 & 43.9\% & ±2.5\% \\
    - & 31 & 0.19 & 0.11 & 43.9\% & ±1.3\% \\
    - & 30 & 0.28 & 0.07 & 43.0\% & ±2.3\% \\
    - & 9 & 0.20 & 0.07 & 37.6\% & ±1.4\% \\
    - & 11 & 0.32 & 0.11 & 37.2\% & ±0.7\% \\
    - & 27 & 0.20 & 0.04 & 34.5\% & ±3.6\% \\
    \citeA{Fuselier2017} & 5 & 0.33 & 0.13 & 33.5\% & ±2.4\% \\
    Chen et al. (2016b) & 1 & 0.37 & 0.10 & 31.0\% & ±1.5\% \\
    - & 22 & 0.58 & 0.10 & 30.9\% & ±1.6\% \\
    - & 16 & 0.46 & 0.15 & 27.9\% & ±2.2\% \\
    \rowcolor{lightgray} \citeA{Burch2016} & 3 & 0.34 & 0.04 & 22.1\% & ±1.5\% \\
    - & 25 & 0.24 & 0.04 & 19.9\% & ±1.6\% \\
    - & 28 & 0.48 & 0.05 & 19.0\% & ±1.6\% \\
    - & 32 & 0.30 & -0.01 & 15.4\% & ±0.8\% \\
    
    \end{tabular}
    \label{tab:Parallel_Table}
\end{table}

\newpage

\subsection{EDR Case Studies}
\label{case_studies}

From Table \ref{tab:Parallel_Table}, we selected three events to investigate further as shown in Figure \ref{fig:EDR_Events} --- the \citeA{Burch2016} event (panel A), the \citeA{Norgren2016} event (panel B), and the \citeA{Phan2016} event (panel C). These events, highlighted in grey, roughly represent the table's low parallel, mid parallel, and high parallel cases respectively and all occur within one week during low guide-field conditions, indicating that these differences are not significantly impacted by large-scale or location based factors.

Starting with the \citeA{Burch2016} event in Figure \ref{fig:EDR_Events} panel A, this event occurred on October 16th, 2015 with a magnetopause crossing over the X-line/dissipation region occurring near 13:07:00 that day as seen in subpanel (a). Using this crossing data, \citeA{Burch2016} estimated the MMS spacecraft's trajectory through the diffusion regions as is shown in subpanel (d). From the estimated trajectory, MMS is believed to have closely crossed the X-line, skimming along the central EDR in an almost vertical cut across the diffusion regions, with the majority of the magnetopause current sheet's crossing (shown in the dashed orange lines) occurring in the EDR and central EDR. This trajectory resulted in the current density time series shown in subpanels (b) and (c), which include a strong -M current structure, and very little resulting parallel current as subpanel (c) highlights with an average percent parallel of $22.1 \pm 1.5\%$. 

Moving to panel B, the \citeA{Norgren2016} event also occurred on October 16th, 2015, with a magnetopause crossing from 10:33:20 to 10:34:00 as shown in subpanel (a). As depicted in subpanel (d), MMS’s trajectory through the diffusion regions is more tilted than the Burch et al. event, with a cut that crosses over both the IDR and EDR, and results in the time series shown in subpanels (a) - (c). These show a strong current density peak in (b), with this current structure becoming primarily parallel in the second half of the magnetopause as is shown in (c). Overall, this gave the Norgren et al. event an average percent parallel of $47.4 \pm 2.9\%$.

Lastly, in panel C, the \citeA{Phan2016} event occurred one week later on October 22nd, 2015 from approximately 06:05:20 and 06:05:30 - subpanel (a) - when the MMS constellation encountered the outer EDR and IDR of the magnetic reconnection event as is seen in subpanel (d). This event's time series in subpanels (a) - (c) is characterized by a strong triple peaked current structure in (b) with the majority of this current becoming parallel to the local magnetic field after 06:05:22.5 as shown in (c). This resulted in a higher average percent parallel of $62.7 \pm 1.6\%$ across its magnetopause crossing. 

Based on these case studies, the parallel current signatures shown in subpanels (c) are significantly tied to how much of MMS's path crossed through the outer EDR and IDR as shown in subpanels (d). This indicates that significant out-of-plane parallel current signatures during a EDR crossing are additional identifying features of the outer-EDR and IDR.

\newpage

\section{2.5D Asymmetric PIC Simulations}

\begin{figure}[htbp]
    \centering
    \includegraphics[width=\textwidth]{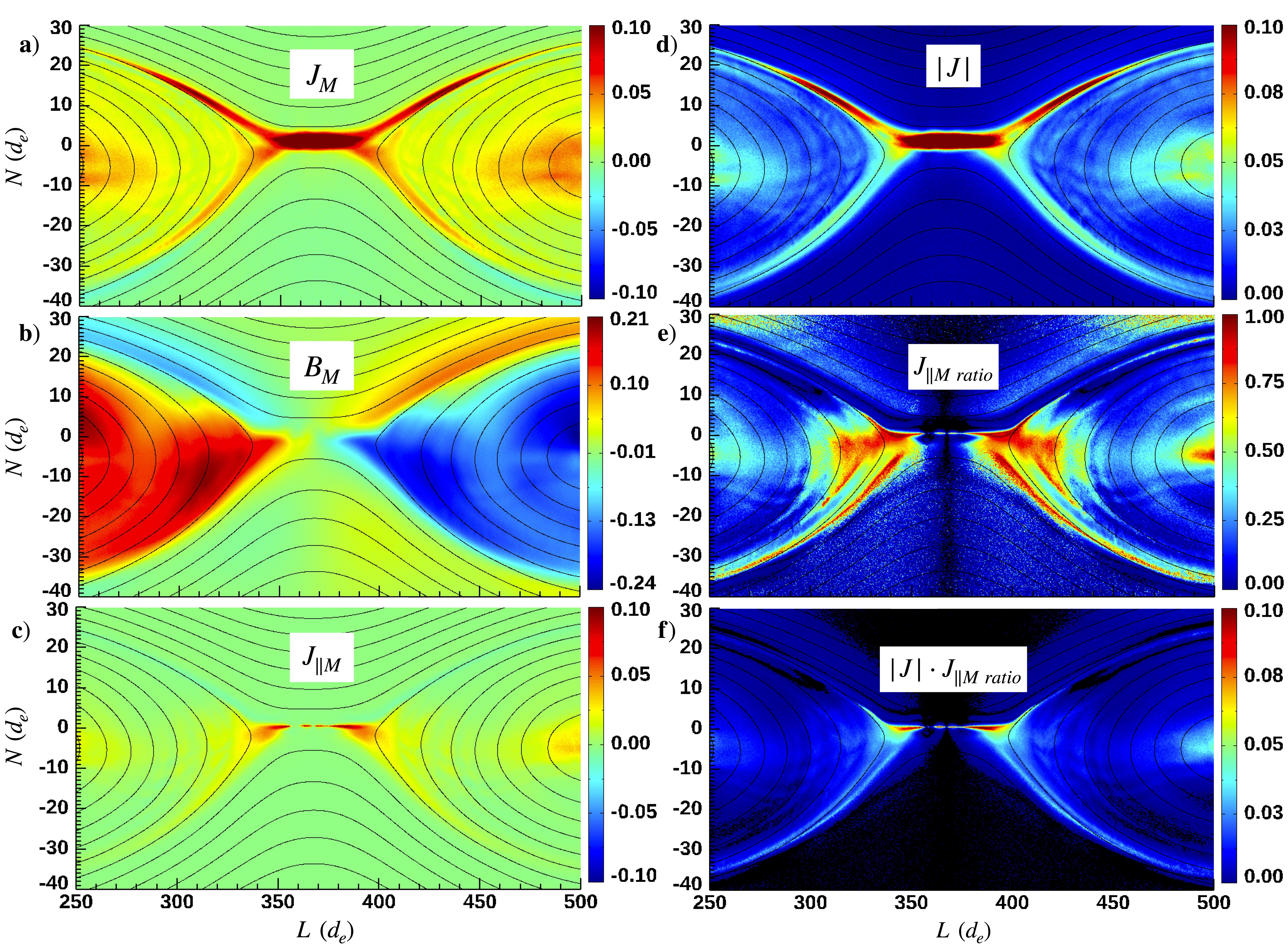}
    \caption{Asymmetric 2.5D PIC simulation results of the current density, Hall magnetic field, and resulting parallel current quantities and ratios across the ion and electron diffusion regions in the reconnection LN plane. a) out-of-plane current density, b) out-of-plane magnetic field, c) out-of-plane parallel current density, d) magnitude of the current density, e) eq. \ref{eq:percent_para} in unit form, f) eq. \ref{eq:percent_para} multiplied by the magnitude of the current density. Note, the M-direction is into the page in the simulation coordinate system, which indicates that panel a)'s current density is in the -M direction, which is consistent with MMS's observations of the reconnection current in Figure \ref{fig:EDR_Events}. Also note the simulation domain is presented in $d_e$, with one $d_i$ being equal to ten $d_e$ in the simulation. The magnetosphere plasma is shown at the top of the diagrams, while the magnetosheath plasma is shown at the bottom.}
    \label{fig:PIC_sim}
\end{figure}

To gain a better understanding of the parallel current MMS encounters in the diffusion regions, and where it arises, we also utilized results from a 2.5D, zero-guide field, asymmetric PIC simulation. This particular simulation used the same initial setup as \citeA{Chen2016_PIC} and starts as an ion-scale current sheet with a domain size of 75 x 25 $d_i$. The mass ratio for this simulation is $m_i/m_e = 100$ and the average number of particles per cell is 3,000 \cite{Chen2016_PIC}. All of the results are taken from $t \omega_{ci} = 68$, which is approximately 8 $\omega_{ci}^{-1}$ after the peak reconnection rate in the same manner as is explained in \citeA{Chen2016_PIC}. Using this simulation setup, we looked at the LN reconnection plane of the diffusion regions, where the M direction is into the page for all simulation results. Magnetic field and current results, and their resulting interaction to create parallel currents, are then shown in Figure \ref{fig:PIC_sim}. 

\newpage

From Figure \ref{fig:PIC_sim}, we can make several observations regarding the current and its parallel component. Panel a) shows the out-of-plane reconnection current $J_M$, which is shown to primarily hug the magnetosphere side of the diffusion regions, with its intensity peaking in the central EDR from approximately 350 to 400 $d_e$. The out-of-plane magnetic field, the Hall magnetic field in this zero-guide-field simulation, is then depicted in panel b). Note how the nominally quadrupolar Hall magnetic field is decidedly bipolar in this case because of the asymmetric boundary conditions with the magnetosheath-side Hall field becoming pulled into the EDR reconnection current sheet. This overlap near the boundary of 350 and 400 $d_e$ then generates the significant out-of-plane parallel current, $J_{\parallel M}$, component seen in panel c), which highlights the outer EDR into the IDR. 

We can also look at the magnitude of the total current as seen in panel d), where, again, the EDR is highlighted by the strong reconnection current. Panel e) then gives us the percent parallel quantity defined from eq. \ref{eq:percent_para}, where we can clearly see how the parallel component of the out-of-plane current becomes increasingly dominant in the IDR and surrounding the central EDR. Note however, that the majority of this high percent parallel occurs in the IDR where the current density is significantly weaker. Thus, we can also multiple this percent parallel quantity with the magnitude of the current, combining the results of panels d) and e), and creating panel f). Looking at the results of panel f), this highlights how, in order to have both a dominant parallel component, and a strong overall current, the IDR and EDR systems must overlap, creating an interaction region that causes a portion of the out-of-plane reconnection current in the outer EDR to become parallel.

\newpage

\section{Discussion}

\subsection{The IDR-EDR Interaction Region, and the Identification of the Outer EDR}

\begin{figure}[htbp]
    \centering
    \includegraphics[width=0.9\textwidth]{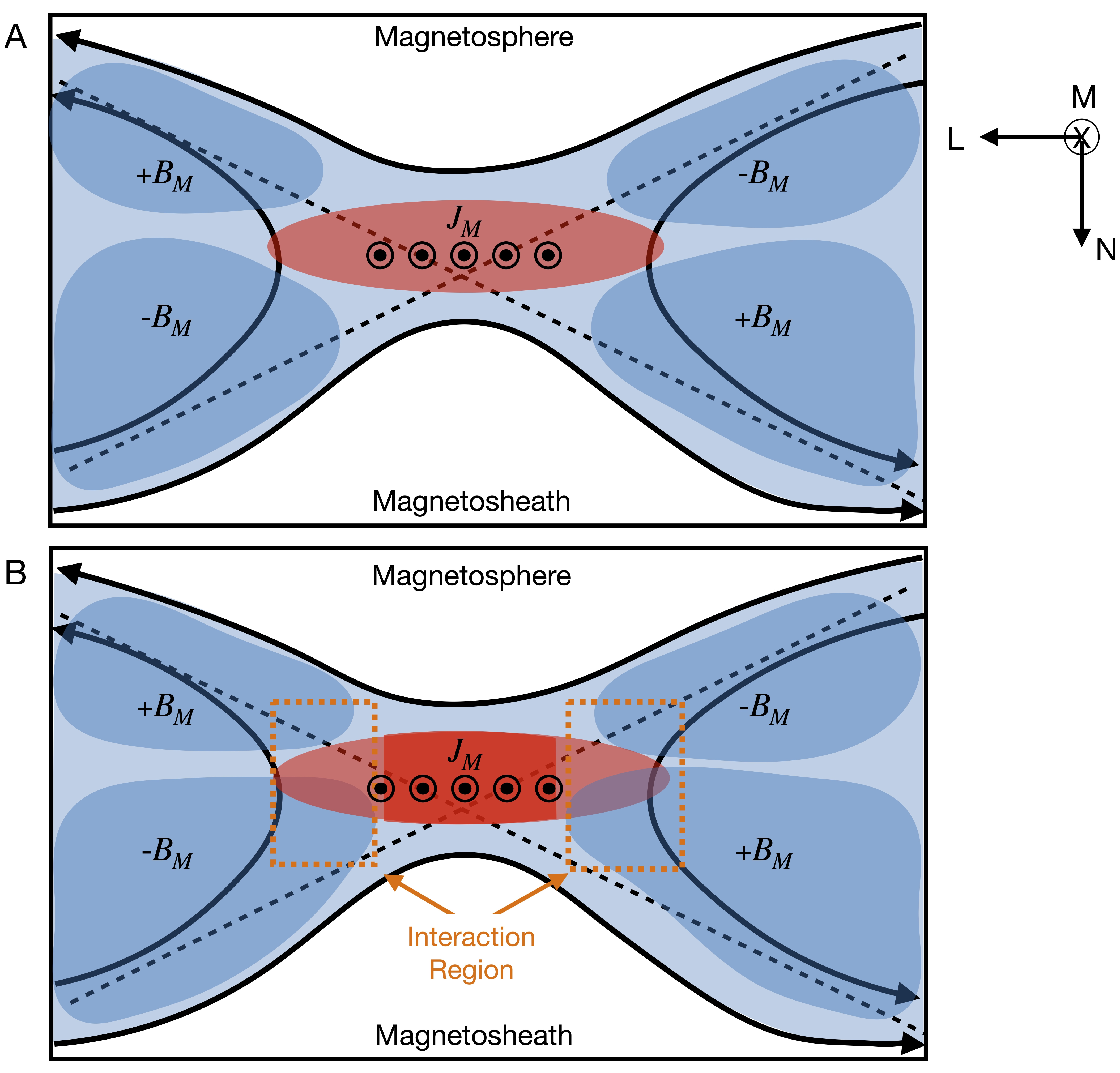}
    \caption{Diagrams of the asymmetric diffusion regions in the reconnection LN plane, with the IDR depicted in blue, and the EDR depicted in red. Panel A represents a traditional diagram of the asymmetric diffusion regions with no overlap between the relevant systems -- the reconnection current sheet in the EDR and the Hall magnetic field in the IDR. Panel B represents our modified view of this diagram which includes the overlap of magnetosheath side of the Hall magnetic field into the reconnection current sheet, which then generates the out-of-plane parallel current discussed in Sections 2 and 3.  The IDR is represented in blue, while the EDR is depicted in red, with the central EDR in dark red, and the outer EDR in light red. The interaction region between the IDR and outer EDR is depicted in the orange dashed regions and helps differentiate between the outer EDR (pale red) and the inner EDR (dark red).}
    \label{fig:diffusion_region_diagram}
\end{figure}

From Figures \ref{fig:EDR_Events} and \ref{fig:PIC_sim}, the interaction of the IDR and EDR systems is apparent and expresses itself in the generation of a significant out-of-plane parallel current in the overlap of the outer EDR with the IDR's asymmetric Hall magnetic field. This interaction is depicted in Figure \ref{fig:diffusion_region_diagram}B by the dashed orange boxes overlapping the reconnection current in the outer EDR (light red) with the dominant magnetosheath-side Hall magnetic field in the IDR. This leaves the central EDR dominated by its perpendicular reconnection current as is shown in Figure \ref{fig:PIC_sim}. Given this distinction, it is highly likely that the out-of-plane parallel current signature could be used to identify the outer EDR and the resulting interaction region in MMS data in the low-guide field regime, thus giving another measure of MMS's location and path through the complex diffusion region system refining our typical diagram of the diffusion regions from Figure 3A to Figure 3B. This also likely applies to symmetric reconnection events in the tail as well, with more investigation into this symmetric case coming in a later study. 

\subsection{Energy Dissipation}

\begin{figure}[htbp]
    \centering
    \includegraphics[width=0.9\textwidth]{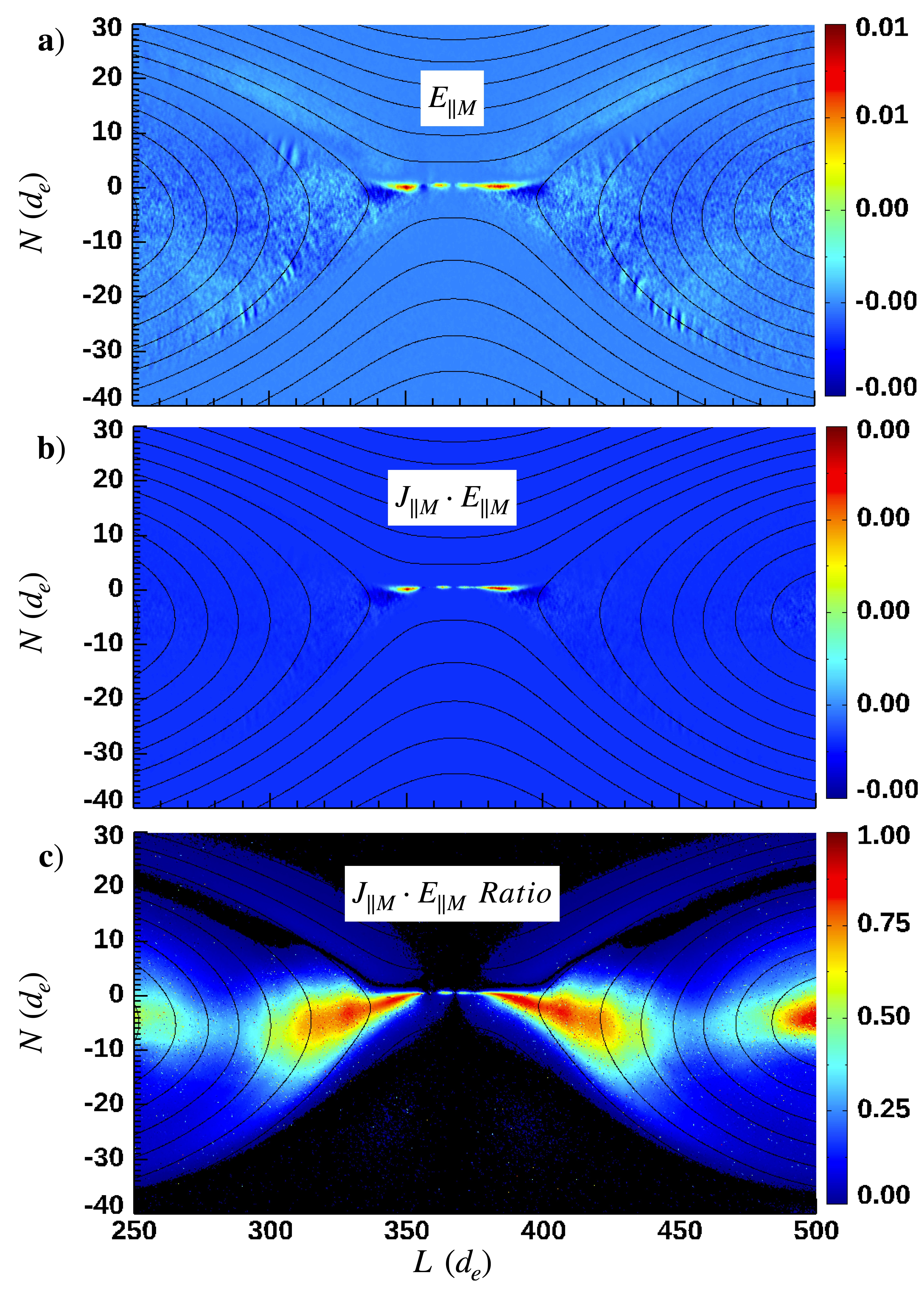}
    \caption{2.5D asymmetric PIC simulation result of the out-of-plane, parallel components of the frame-independent dissipation measure $J_\parallel E_\parallel$ -- e.g. \citeA{Zenitani2011}. a) out-of-plane parallel electric field, b) the out-of-plane component of the parallel dissipation, c) the ratio contribution of the out-of-plane component of the parallel dissipation to the total parallel dissipation, in the same method as eq. \ref{eq:percent_para}.}
    \label{fig:PIC_sim_E}
\end{figure}

The out-of-plane parallel current structure also has implications for energy dissipation in the outer EDR region. Specifically, given \citeA{Zenitani2011}'s definition of the $J \cdot E'$ term, and its decomposition into non-negligible parallel and perpendicular components via \citeA{Shay2016}, we have $D_e = E_\parallel J_\parallel + \textbf{J}_\perp \cdot(\textbf{E} + \textbf{V}_e \times \textbf{B})$. Note, in \citeA{Shay2016}, the parallel component of $D_e$, $J_\parallel E_\parallel$ is found as having a large positive value near the X-line, which ultimately cancels with the perpendicular term to give a near zero total value for $D_e$. However, where this positive $J_\parallel E_\parallel$ comes from is not discussed, and ultimately relies on the overlap of both a significant parallel E-field and parallel current density component. As previously shown in Figure \ref{fig:PIC_sim}, the out-of-plane parallel current density can provide this significant parallel contribution, especially in the outer EDR. This combined with the out-of-plane E-field component shown in Figure \ref{fig:PIC_sim_E} panel a), then gives rise to the out-of-plane parallel energy dissipation factor shown in panel b), which indicates a measurable energy contribution from the overlap of these two elements in the outer EDR. Additionally, if we construct a ratio factor in the same method as eq. \ref{eq:percent_para} for the $J_{\parallel M} \cdot E_{\parallel M}$ component, comparing its contribution to the total $J_\parallel E_\parallel$, we can see in panel c) that this out-of-plane contribution is the dominant contributor to the energy dissipation in the outer EDR and into the IDR. This indicates that the interaction region between the IDR and outer EDR not only gives rise to a significant out-of-plane current density in this region, but is also responsible for contributions to the region's energy dissipation signature as well.

\newpage

\section{Summary and Conclusions}

In this paper, we compared statistical results from 26 no-to-low guide-field dayside EDR crossings with a 2.5D asymmetric PIC simulation of the diffusion regions. We first selected three of the 26 EDR events and analysed them as case studies of the low, medium, and high out-of-plane parallel current contribution across the magnetopause current sheet. The parallel current contribution found in each of the three events corresponded to how the spacecraft crossed through the diffusion regions, with the more the spacecraft encountered the outer EDR and IDR leading to more of the total current becoming parallel to the local magnetic field. This implies that the out-of-plane parallel current signature is tied to an interaction region between the outer EDR's current sheet and the IDR's Hall magnetic field, and could act as an indicator of the outer EDR region in magnetopause crossing data.

We then utilized the simulation domain of a 2.5D asymmetric PIC simulation to see how the out-of-plane current structure of the EDR and IDR should appear. The simulated results echoed the case study findings by showing how the out-of-plane parallel current contribution is highly tied to the IDR and the interaction region between the IDR and outer EDR. Additionally, we found that the resulting out-of-plane current density could not only be used as an additional identifying signature of the outer EDR, but is also responsible for a significant portion of the parallel energy dissipation in this outer EDR region.

Overall, these findings indicate that significant out-of-plane parallel current signatures and contributions are intrinsically tied to the outer EDR and interaction region with the IDR, distinguishing these regions from the central EDR in both asymmetric magnetopause case studies and simulated results. Investigation into this parallel current signature in symmetric diffusion regions, and higher guide-field cases, as well as electron-only reconnection is ongoing. 

\section{Open Research}
The MMS data used in this study is publicly available from the FPI and FIELDS datasets provided at the MMS Science Data Center, Laboratory for Atmospheric and Space Physics (LASP), University of Colorado Boulder \cite{MMS_dataset} as well as from the following MMS1 datasets - \citeA{MMS1FPI_i,MMS1FPI_e,MMS1FIE}. The averaged MMS crossing data presented in Table \ref{tab:Parallel_Table} is available through a Harvard Dataverse public database \cite{BeedleData2024}. 

\acknowledgments
We thank Li-Jen Chen for the use of the PIC simulation setup used in this study. We also thank the entire MMS team and instrument leads for the data access and support. Additionally, we thank the pySPEDAS team for their support and data analysis tools. This research was supported by the NASA Magnetospheric Multiscale Mission in association with NASA contract NNG04EB99C. J. M. H. B. and V. M. U. were supported through the cooperative agreement 80NSSC21M0180. J. M. H. B was also supported by the NASA MMS grant 14N820-UZSPRT.


%
%




\bibliography{main.bib}

%
%
%
%
%

\end{document}